\documentclass[showpacs,preprintnumbers,twocolumn,amsmath,amssymb,
groupedaddress,superscriptaddress]{revtex4}
\usepackage{graphicx}
\usepackage{dcolumn}
\usepackage{bm}

\begin{document}

\title{Study of $^{6}$He+$^{12}$C Elastic Scattering Using a
Microscopic Optical Potential}

\author{V.~K.~Lukyanov}
\affiliation{Joint Institute for Nuclear Research, Dubna 141980,
Russia}

\author{D.~N.~Kadrev}
\affiliation{Institute for Nuclear Research and Nuclear Energy,
Bulgarian Academy of Sciences, Sofia 1784, Bulgaria}

\author{E.~V.~Zemlyanaya}
\affiliation{Joint Institute for Nuclear Research, Dubna 141980,
Russia}

\author{A.~N.~Antonov}
\affiliation{Institute for Nuclear Research and Nuclear Energy,
Bulgarian Academy of Sciences, Sofia 1784, Bulgaria}

\author{K.~V.~Lukyanov}
\affiliation{Joint Institute for Nuclear Research, Dubna 141980,
Russia}

\author{M.~K.~Gaidarov}
\affiliation{Institute for Nuclear Research and Nuclear Energy,
Bulgarian Academy of Sciences, Sofia 1784, Bulgaria}


\begin{abstract}
The $^6$He+$^{12}$C elastic scattering data at beam energies of 3,
38.3 and 41.6 MeV/nucleon  are studied utilizing the microscopic
optical potentials obtained by a double-folding procedure and also
by using those inherent in the high-energy approximation. The
calculated optical potentials are based on the neutron and proton
density distributions of colliding nuclei established in an
appropriate model for $^6$He and obtained from the electron
scattering form factors for $^{12}$C. The depths of the real and
imaginary parts of the microscopic optical potentials are
considered as fitting parameters. At low energy the volume optical
potentials reproduce sufficiently well the experimental data. At
higher energies, generally, additional surface terms having form
of a derivative of the imaginary part of the microscopic optical
potential are needed. The problem of ambiguity of adjusted optical
potentials is resolved requiring the respective volume integrals
to obey the determined dependence on the collision energy.
Estimations of the Pauli blocking effects on the optical
potentials and cross sections are also given and discussed.
Conclusions on the role of the aforesaid effects and on the
mechanism of the considered processes are made.
\end{abstract}


\maketitle

\section{Introduction\label{s:intro}}

Experimental and theoretical studies of exotic light nuclei with a
localized nuclear core and a dilute few-neutron halo or skin have
been an important and advanced area in the nuclear physics in the
past decades. The availability of radioactive ion beams facilities
made it possible to carry out many experiments and to get more
information regarding the structure of these nuclei and the
respective reaction mechanisms (for more information see, e.g., the
recent review of the problem in Ref.~\cite{Keeley2009}). In this
sense, $^{6}$He is a typical nucleus having the weak binding energy
and extended neutron halo in its periphery. The latter is the reason
why in collisions with the proton and nuclear targets the projectile
nucleus $^{6}$He is breaking up with a comparably large probability
that causes the flux loss in the elastic channel. Therefore, the
study of elastic scattering of $^{6}$He on protons or light targets
is a powerful tool to get information on peculiarities of the
mechanism of such processes.

The data on cross sections of processes  with light exotic nuclei
have been analyzed using various phenomenological and microscopic
methods. Among the latter we should mention the microscopic
analysis using the coordinate-space g-matrix folding method (e.g.,
Ref.~\cite{Amos2005} and references therein), as well as works
where the real part of the optical potential (ReOP) is
microscopically calculated (e.g., Ref.~\cite{Avrigeanu2000}) using
the folding approach (e.g., Refs.~
\cite{Satchler79,Khoa1993,Khoa2000,Karataglidis2000}). Usually the
imaginary part  of the OP's (ImOP) and the spin-orbit terms have
been determined phenomenologically that has led to the usage of a
number of fitting parameters. In our previous works
\cite{Lukyanov2007,Lukyanov2009} instead of using a
phenomenological imaginary part of OP we have performed
calculations of $^{6}$He$+p$ \cite{Lukyanov2007} and $^{8}$He$+p$
\cite{Lukyanov2009} elastic differential cross sections by means
of the microscopic OP with the imaginary part taken from the OP
derived in \cite{Lukyanov2004,Shukla2003} in the frameworks of the
high-energy approximation (HEA) \cite{Glauber,Sitenko,Czyz1969}
that is known as the Glauber theory. In the case of $^{6}$He$+p$
elastic scattering, it has been shown in our previous study
\cite{Lukyanov2007} that the depth of the imaginary part of the
respective microscopically calculated OP has to be appreciably
changed to get an agreement with the existing experimental data.
The present study of $^{6}$He+$^{12}$C scattering could give a
novel information on mechanism of the process due to the more
complicated dependence of the microscopic OP not only on the
density of the projectile $^{6}$He but also on the density of the
target nucleus.

In the last years a number of works has been devoted also to the
elastic scattering of $^{6}$He on $^{12}$C nucleus and,
particularly, to the study of the mechanism of this process
including the role of breakup channels. In the present paper we
perform an analysis of the $^{6}$He+$^{12}$C elastic scattering
data at three beam energies $E=3$ \cite{Milin2004}, 38.3
\cite{Lapoux2002} and 41.6 MeV/nucleon \cite{Khalili96} using the
microscopically calculated OP. The data have been already
considered individually in the frameworks of other theoretical
models. In Ref.~\cite{Milin2004} the differential cross sections
at total energy 18 MeV were analyzed by means of Woods-Saxon (WS)
OP with radius parameter of the imaginary part $R_I=5.38$ fm being
about twice larger than that of the real part ($R_R=2.4-2.7$ fm).
The so large absorbtion radius in elastic channel may thought to
be caused by breakup channels which take place at the far
periphery of OP. Recently, the same $^{6}$He+$^{12}$C elastic
scattering data were fitted in Ref.~\cite{Kucuk2009} by OP having
a squared WS real part and a standard WS shape for its imaginary
part. Furthermore,  a larger radius $R_I=6.17$ fm was obtained for
the latter. The different forms of the real part of OP and also
the values of $R_I$ in Refs.~\cite{Milin2004} and \cite{Kucuk2009}
just reflect the known problem of the ambiguity of parameters of
phenomenological OP's when fitting them to the restricted amount
of experimental data. In principle, this problem does not arise in
microscopic OP's whose basic parameters have been already
established by fitting to other data. In this line, a part of the
problem was overcome in Ref.~\cite{Lapoux2002}, where the elastic
scattering data at $E=38.3$ MeV/nucleon were analyzed using the
semi-microscopic OP. Its real part was defined in a double-folding
model including the direct and exchange convolution integrals,
while the imaginary part was taken phenomenologically in the WS
form. Adjusting the latter to the data at relatively small angles,
the reduced radius and diffuseness parameters were obtained in the
range of $r_{I}=1.471-1.569$ fm and $a_I=0.358-0.524$ fm. Then, to
get a better agreement at larger angles, the dynamic polarization
potential (DPP) in the form of a derivative of a WS function was
added to the volume OP. It affects strongly the total OP in the
surface region at radii around $4-5$ fm.

Recently, in Ref.~\cite{El-Azab2008} five Gaussian-like forms for
the $^{6}$He matter density distributions were tested using the
real part of OP without the exchange term \cite{Satchler79}
together with the 5-parameter phenomenological imaginary part
(volume and surface) based on the WS form. This model was adopted
to study the data at 38.3 and 41.6 MeV/nucleon. Also, the authors
generated another (microscopic) OP by involving in folding
calculations the complex Jeukenne-Lejeunne-Mahaux (JLM) effective
nucleon-nucleon ($NN$) potential \cite{Jeukenne}. Doing so, the
optimized values of the fitted parameters were fixed and then, to
improve slightly the fits to data the repulsive real DPP term with
two free parameters was introduced, as well. In this way, the
ambiguity problem is retained just in such semi-microscopic models
of OP.

In Ref.~\cite{Abu-Ibrahim2003}, an attention was paid to the
dynamic polarization term in OP. A Monte Carlo method was
developed to calculate both the $^{6}$He variational wave function
and the Glauber amplitude of the microscopic scattering of the
nucleons of $^{6}$He by the nucleus $^{12}$C as a whole. To this
end the phenomenological $p+$$^{12}$C OP was utilized and, as a
result, the full $^{6}$He+$^{12}$C OP was restored from the
calculated Glauber eikonal phase. Then, the difference between
this OP and the single-folding OP estimated without accounting for
the Glauber multiple-scattering terms, was defined as DPP
responsible for the breakup channels. One can see from Fig.~5 of
Ref.~\cite{Abu-Ibrahim2003} that at 40 MeV/nucleon in the range of
$4-5$ fm  the imaginary single-folding potential (negative) $W$ is
about $11-3$ MeV and the DPP makes it deeper by about $7-3$ MeV,
so the full eikonal $W$ has to be about $18-6$ MeV. The certain
way to investigate simultaneously the effects of elastic and
breakup processes can be performed in the framework of
coupled-channel (CC) methods. In
Refs.~\cite{Khalili96,Matsumoto2009} these methods have been
elaborated to estimate the effects of the elastic and inelastic
breakups of the projectile nucleus $^{6}$He$\rightarrow
\alpha$+$n$+$n$ on the elastic scattering of $^{6}$He+$^{12}$C.
Two models for the $^{6}$He structure were explored in
Ref.~\cite{Khalili96}. One of them uses the modified wave function
from \cite{Zhukov1993} of the 3-body $\alpha$+$n$+$n$ system
considered as a ``Borromean'' one. The second model constructs the
$\alpha$- ``di-neutron'' potential and the corresponding 2-body
wave function. So far the input potentials $U_{\alpha+n}$,
$U_{\alpha+2n}$ and $U_{\alpha+ ^{12}C}$, $U_{n+ ^{12}C}$, $U_{2n+
^{12}C}$ have been taken from the respective fits and estimations.
These both models are completely parameter-free and explain the
elastic scattering data at 41.6 MeV/nucleon fairly well. The
analysis of the breakup effects of $^{6}$He on the elastic
scattering is done in Ref.~\cite{Matsumoto2009} using the
so-called continuum-discretized CC method. The reaction system is
described as a four-body model of $\alpha$+$n$+$n$+$^{12}$C. The
wave function of the bound and continuum states of the 3-body
system $\alpha$+$n$+$n$ is presented applying the specified
Gaussian expansion method, and the ground state wave function of
$^{12}$C is calculated by the microscopic 3$\alpha$ cluster model.
The resulting microscopic OP was calculated as double-folding
integral with the calculated densities of $^{6}$He and $^{12}$C,
and then multiplied by the complex factor $(N_{R}+iN_{I})$ with
coefficients optimized by a fit to the elastic scattering data.
The exchange terms in the folding OP were neglected. The results
for the $^{6}$He+$^{12}$C scattering showed that the optimum value
of $N_{I}$ is 0.5 at 3 MeV/nucleon and 0.3 at 38.3 MeV/nucleon,
respectively, while $N_{R}=1$. Also, it was shown that the effects
of the coupling of channels are more important at comparatively
large angles. The coupling smooths the diffraction structure of
the differential cross sections at $E=3$ MeV/nucleon and shifts
down the corresponding curve at $E=38.3$ MeV/nucleon calculated
without coupling.

It can be mentioned that calculations by coupled reaction channel
models with accounting for the cluster and continuum states are
encouraged to study their sensitivity to the input information on
the reaction mechanism. On the other hand, the breakup reactions
reveal themselves through the dynamic polarization terms in the
full OP for elastic scattering. The explicit information on these
channels can be obtained from the unambiguous OP obtained from the
respective analysis of the elastic scattering experimental data.
In our paper we try to realize the following idea. We start
analyzing the elastic scattering data by the microscopic optical
potential obtained in Ref.~\cite{Lukyanov2004}. Its real part
includes the direct and exchange terms that are the same used in
Ref.~\cite{Lapoux2002}. The imaginary part of OP is based on the
Glauber theory of high-energy scattering of complex systems and is
an integral which folds the nucleon-nucleon scattering amplitude
$f_{NN}$ with the density distribution functions of the bare
nucleons of colliding nuclei. Therefore, first, this OP consists
of parameter-free real and imaginary parts defined by the
respective terms of $f_{NN}$. However, this potential reveals only
the single-particle physical nature of the colliding nuclei
because it depends on the single-particle nuclear densities. Then,
we assume that additional terms to our basic OP (which have to be
added to explain the experimental data) may be considered as a
consequence of the presence of more complicated channels. In the
case of the loosely bound $^{6}$He projectile these terms are
thought to arise due to the breakup channels. Thus the main effort
should be directed to minimize the ambiguities in the fitted OP's
by studying differential elastic cross sections at different
energies and to involve external physical conditions in order to
make as narrow as possible the corridor of the deviations of
selected theoretical curves.

The paper is organized as follows. The theoretical scheme to
calculate the real and imaginary parts of the OP is given in
Sec.~\ref{s:theory}. The results of the calculations of OP's and the
$^{6}$He+$^{12}$C elastic scattering differential cross sections and
their discussion are presented in Sec.~\ref{s:results}. The summary
of the work and conclusions are given in Sec.~\ref{s:conclusions}.

\section{The microscopic optical potential\label{s:theory}}

Here we give  the main expressions for the real and imaginary parts
of the nucleus-nucleus OP
\begin{equation}\label{eq:1}
U(r) = V^{DF}(r) + iW(r).
\end{equation}
The real part $V^{DF}$ consists of the direct and exchange
double-folding (DF) integrals that include an effective $NN$
potential and  density distribution functions of colliding nuclei:
\begin{equation}\label{eq:2}
V^{DF}(r)= V^D(r) + V^{EX}(r).
\end{equation}
The formalism of the DF potentials is described in details, e.g., in
Refs.~\cite{Satchler79,Khoa2000}. In general, in Eq.~(\ref{eq:2})
$V^{D}$ and $V^{EX}$ are composed from the isoscalar (IS) and
isovector (IV) contributions, but in the considered case the
isovector part is omitted because $Z=N$ in the target nucleus
$^{12}$C and, thus, one can write:
\begin{equation}\label{eq:3}
V^D(r) = \int d^3 r_p d^3 r_t  {\rho}_p({\bf r}_p) {\rho}_t ({\bf
r}_t) v_{NN}^D(s),
\end{equation}
\begin{eqnarray}
V^{EX}(r)&=& \int d^3 r_p d^3 r_t   {\rho}_p({\bf r}_p, {\bf
r}_p+ {\bf s})  {\rho}_t({\bf r}_t, {\bf r}_t-{\bf s})
\nonumber \\
& & \times  v_{NN}^{EX}(s)  \exp\left[ \frac{i{\bf K}(r)\cdot
s}{M}\right], \label{eq:4}
\end{eqnarray}
where ${\bf s}={\bf r}+{\bf r}_t-{\bf r}_p$ is the vector between two
nucleons, one of which belongs to the projectile and another one to
the target nucleus. In Eq.~(\ref{eq:3}) $\rho_p({\bf r}_p)$ and
$\rho_t({\bf r}_t)$ are the densities of the projectile and the
target, respectively. In Eq.~(\ref{eq:4}) $\rho_p({\bf r}_p, {\bf
r}_p+{\bf s})$ and $\rho_t({\bf r}_t, {\bf r}_t-{\bf s})$ are the
density matrices for the projectile and the target that are usually
taken in an approximate form \cite{Negele72}. In the modern
calculations of the DF potentials the effective interaction
$v_{NN}^{D}$ (of CDM3Y6-type) based on the Paris $NN$ forces and
having the form
\begin{equation}\label{eq:5}
v_{NN}^{D}(E,\rho, s)=g(E) F(\rho) v(s)
\end{equation}
is usually applied with the distance dependence given by
\begin{equation}\label{eq:6}
v(s)=\sum_{i=1}^{3} N_i \frac{\exp(-\mu_i s)}{\mu_i s},
\end{equation}
and with terms of the energy and density dependencies:
\begin{equation}\label{eq:7}
g(E)=1-0.003E, \quad F(\rho)=C\Bigl [1+\alpha
e^{-\beta\rho}-\gamma\rho\Bigr].
\end{equation}
The energy dependent factor in Eq.~(\ref{eq:7}) is taken to be a
linear function of the bombarding energy per nucleon, while $\rho$
in $F(\rho)$ is the sum of the projectile and target densities,
$\rho=\rho_{p}+\rho_{t}$. The parameters $N_i,\,\mu_i$
[Eq.~(\ref{eq:6})], $C,\,\alpha,\,\beta,\,\gamma$
[Eq.~(\ref{eq:7})], and all details of the mathematical treatments
and calculations are given in Refs.~\cite{Khoa2000,Kostya2007}.

In Eq.~(\ref{eq:4}) $v_{NN}^{EX}$ is the exchange part of the
effective $NN$ interaction.  It is important to note that the
energy dependence of $V^{EX}$ arises primarily from the
contribution of the exponent in the integrand of Eq.~(\ref{eq:4}).
Indeed, there the local nucleus-nucleus momentum
\begin{equation}\label{eq:8}
K(r)=\left \{\frac{2Mm}{\hbar^2}\left[E-V^{DF}(r)-V_c(r)\right
]\right \}^{1/2}
\end{equation}
with $A_{p}$, $A_{t}$, $m$ being the projectile and target atomic
numbers and the nucleon mass, and $M=A_pA_t/(A_p+A_t)$. As can be
seen, $K(r)$ depends on the folding potential $V^{DF}(r)$ that has
to be calculated itself and, therefore, we have to deal with a
typical non-linear problem. Usually, two different kinds of
effective $NN$ potentials are employed in calculations, namely the
Paris CDM3Y6 and the Reid DDM3Y1 $NN$ interactions, which are
defined by two different sets of the aforementioned parameters.
The direct parts of these potentials have different signs and, for
example, in the case of CDM3Y6 forces the $V^{EX}$ is negative
while $V^{D}$ is positive. So, if in the calculations one takes
only the direct part of $V^{DF}$ with the Paris M3Y $NN$ forces,
then the corresponding real part of such OP is positive one.
Therefore, one should proceed carefully when neglecting the
exchange part of OP.

Concerning the imaginary part of our OP, we take it in two forms.
In the first case the imaginary part has the  same form as the
real one but with different strength. At the same time we test
another shape of the imaginary part that corresponds to the full
microscopic OP derived in Refs.~\cite{Lukyanov2004,Shukla2003}
within the HEA \cite{Glauber,Sitenko}. In the momentum
representation this OP has the form
\begin{eqnarray}\label{eq:9}
U^H_{opt}(r)&=&-\frac{E}{k}{\bar\sigma}_{N} (i + {\bar\alpha}_{N})\frac{1}{(2\pi)^3} \nonumber \\
& & \times \int e^{\displaystyle{-i\bf q \bf r}}{\rho}_p(q){\rho}_t(q)f_N(q) d^3q.
\end{eqnarray}
Here $\bar\sigma_{N}$ and $\bar\alpha_{N}$ are the averaged over
isospins of nuclei the $NN$ total scattering cross section and the
ratio of real to imaginary parts of the forward $NN$ amplitude,
both being parameterized, e.g., in Refs.~\cite{Charagi92,
Shukla2001}. The $NN$ form factor is taken as
$f_N(q)=\exp(-q^2\beta/2)$ with the slope parameter $\beta=0.219$
fm$^2$ \cite{Alkhazov1978}. It is easy to verify that the eikonal
phase for this potential
\begin{eqnarray}\label{eq:10}
\Phi(b)=-\frac{1}{\hbar v}\int_{-\infty}^\infty
U^H_{opt}(\sqrt{b^{2}+z^{2}})dz
\end{eqnarray}
is reduced to the HEA microscopic phase $\Phi_N$ for
nucleus-nucleus scattering. To this end, let us substitute
(\ref{eq:9}) in (\ref{eq:10}) and simplify the latter in the
cylindrical coordinate system  where $d^3q=q_{\perp}dq_{\perp}
d\phi dq_{||}$ and ${\bf q}{\bf r}= q_{\perp}b\cos\phi+q_{||}z$, ~
$q^2=q_{\perp}^2+q_{||}^2,~q_{\perp}=q\cos(\vartheta/2), ~q_{||}=
q\sin(\vartheta/2)$. Here $\vartheta$ is the angle of scattering,
$b$ is the impact parameter of the projectile trajectory of motion
directed straight ahead along the $z$ axis.  In HEA one neglects
the longitudinal part of the transfer momentum $q_{||}\ll
q_{\perp}$ and the small angle terms  $q_{||}z\simeq
q_{||}2R\simeq kR\vartheta^2\ll 1$ to give $q\simeq q_{\perp}$ at
$\vartheta\ll\sqrt{1/kR}$. Therefore, the smooth functions become
$\rho(q)\to\rho(q_\perp)$, and $f_N(q)\to f_N(q_{\perp})$, and
then one can perform integrations $\int_{-\infty}^\infty
dq_{||}\exp(-iq_{||}z)=2\pi\delta(z)$ and $\int_0^{2\pi}
d\phi\exp(-iq_{\perp}b\cos\phi)=2\pi J_0(q_{\perp}b)$. Thus, as a
result we obtain the standard form of the HEA phase in the
so-called optical limit of the Glauber theory:
\begin{eqnarray}\label{eq:11}
\Phi_N(b)&=& \frac{1}{4\pi}{\bar\sigma}_{N}(i+{\bar\alpha}_{N})\nonumber\\
 & & \times \int_0^\infty
J_0(qb){\rho}_p(q){\rho}_t(q)f_N(q)\,qdq.
\end{eqnarray}
Here one sets $q_\perp=q=2k\sin(\vartheta/2)$ that is valid at small
angles of scattering $\vartheta\ll\sqrt{1/kR}$.

As a rule this phase is employed to estimate the HEA elastic
scattering amplitude
\begin{equation}\label{eq:12}
f(\vartheta) = ik \int_{0}^{\infty} ~J_{0}(qb) \Bigl[1 -
e^{\displaystyle{i\Phi_N(b)}}\Bigr] bdb.
\end{equation}

In applications of $f(\vartheta)$ from Eq.~(\ref{eq:12}) the main
limitations $E \gg |U|$ and $\vartheta\ll\sqrt{1/kR}$ are
connected with the basic assumption that the integration in
Eqs.~(\ref{eq:10})-(\ref{eq:12}) is performed along the $z$-axis
with a straight-line classical trajectory of motion. To correct
partly this approximation at lower  energies and larger angles one
can take into account the trajectory distortion. For the Coulomb
distortion the respective prescription was used, e.g., in
Ref.~\cite{Vitturi1987}, where the impact parameter $b$ in
(\ref{eq:12}) was replaced by $b_c=\bar a+\sqrt{{\bar a}^2+b^2}$
with $\bar a=Z_1Z_2e^2/2E_{c.m.}$ being the half-distance of
closest approach in the Coulomb field of a point charge. The
approximated method for the inclusion of a distortion in presence
of a nuclear potential was formulated, e.g., in
Ref.~\cite{Brink1981}. However, all these attempts can not fully
account for distortion effects of the classical trajectory caused
by the total complex OP. Instead, the conventional way to resolve
this problem is to compute numerically the Schr\"{o}dinger
equation with the initial OP given by Eq.~(\ref{eq:9}). Moreover,
when using this original OP in the wave equation, it becomes not
necessary to neglect the longitudinal terms in the momentum
transfer. In this way, one expects that the initial OP
(\ref{eq:9}) can be used not only for the scattering at high
energies but also for comparably lower energies and for wider
range of scattering angles.

Hereafter we shall use only the imaginary part of the full OP
(\ref{eq:9}) transformed (by using the equality $\int d\Omega_q
\exp{\displaystyle (-i{\bf q}{\bf r})}=4\pi j_0(qr)$) to the form
\begin{eqnarray}\label{eq:13}
W^H(r) &=& -\frac{1}{2\pi^2}\frac{E}{k}{\bar\sigma}_{N} \nonumber \\
& & \times \int_0^{\infty} j_0(qr){\rho}_p(q){\rho}_t(q){f}_N(q) q^2dq.
\end{eqnarray}

In the further calculations the microscopic volume optical
potential has the following form:
\begin{equation}\label{eq:14}
U(r) = N_R V^{DF}(r) + i N_I W(r),
\end{equation}
where $W(r)$ is taken to be equal either to $V^{DF}(r)$ or to
$W^{H}(r)$. The parameters $N_R$ and $N_I$ entering Eq.~(\ref{eq:14})
renormalize the strength of OP and are fitted by comparison with the
experimental cross sections. In the present work, attempting to
simulate the surface effects caused by the polarization potential
\cite{Lapoux2001,Khoa95,Hussein94}, we add to the volume potential
[Eq.~(\ref{eq:14})] the respective surface terms. Usually, they are
taken as a derivative of the imaginary part of OP, as follows:
\begin{eqnarray}
W^{sf}(r)& =& -i N^{sf}_I\frac{d W(r)}{dr},\label{eq:15}\\
         & =& -i N^{sf}_Ir \frac{d W(r)}{dr},\label{eq:16}\\
         & =& -i N^{sf}_Ir^2 \frac{d W(r)}{dr},\label{eq:17}\\
         & =& -i N^{sf}_I\frac{d W(r-\delta)}{dr},\label{eq:18}
\end{eqnarray}
where $N_{I}^{sf}$ is also a fitting parameter, $\delta$ gives the
shift of the potential (\ref{eq:18}) and in our case is fixed to
be $\delta=1$ fm.

Concluding this section, we would like to emphasize that our basic
OP contains the same real part as that one in
Ref.~\cite{Lapoux2002}, but instead of the phenomenological ImOP,
we utilize a microscopically calculated imaginary potential. In
addition, in contrast to Ref.~\cite{El-Azab2008} where only the
direct part of the ReOP and a part of the ImOP were  calculated
microscopically, in the present work we include also microscopic
exchange part of the OP. Concerning the comparison with the
experimental data, we consider not only selected data (as e.g., at
$E=38.3$ MeV/nucleon in \cite{Lapoux2002}, and at $E=38.3$ and
41.6 MeV/nucleon in \cite{El-Azab2008}) but add also the data at
fairly low energy $E=3$ MeV/nucleon and analyze simultaneously the
three sets of data. This allows us to determine the ambiguities
when adjusting the values of the OP parameters
[Eqs.~(\ref{eq:14})-(\ref{eq:18})] because  we include an
additional condition in the fitting procedure. Namely, we consider
also the energy behavior of the volume integrals of ReOP and ImOP
as an additional physical constraint. In this way, the information
on the dynamical polarization potentials obtained from such more
precise analysis can be considered as more reliable.

\section{Results and discussion\label{s:results}}

In this section we present the results of the calculations of the
microscopic OP's and the respective elastic scattering differential
cross sections at energies $E < 100$ MeV/nucleon obtained following
the theoretical scheme in Sec.~\ref{s:theory}. In contrast to the
cases of $^{6,8}$He$+p$ elastic scattering where only the density of
the projectile $^{6,8}$He takes part in the calculations, in our case
both densities, of the projectile $^{6}$He and the target $^{12}$C,
have to be included in the calculations of the OP (see
Eqs.~(\ref{eq:3}) and (\ref{eq:4})). The results of our work
\cite{Lukyanov2007} on the $^{6}$He$+p$ elastic scattering  at $E <
100$ MeV/nucleon showed that the large-scale shell model (LSSM)
density of $^{6}$He microscopically calculated in
Ref.~\cite{Karataglidis2000} using a complete 4$\hbar\omega$
shell-model space and the Woods-Saxon single-particle wave function
basis with realistic exponential asymptotic behavior is the most
preferable one and it is used also in the present work. For $^{12}$C
we use the symmetrized Fermi-type density with the radius and
diffuseness parameters $c=3.593$ fm and $a=0.493$ fm from
Ref.~\cite{LukyanovZemSlow2004}. They were obtained by defolding the
$^{12}$C charge density distribution deduced in
Ref.~\cite{BurLuk1977} from analysis of the corresponding electron
scattering form factors.

\begin{figure}
\includegraphics[width=1.0\linewidth]{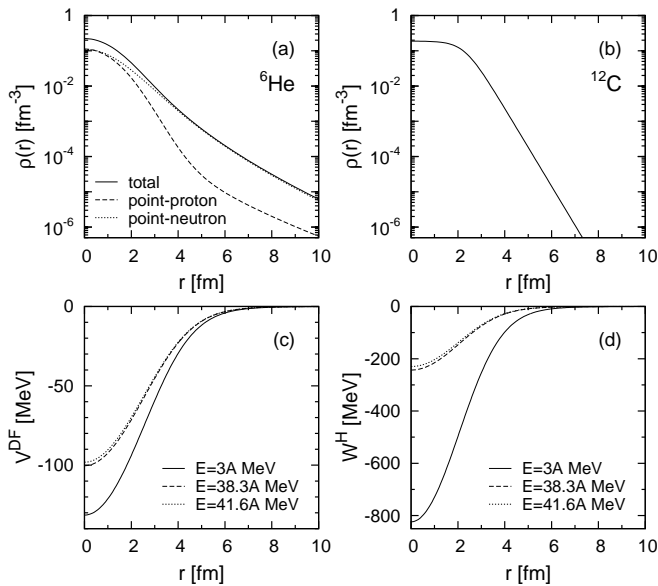}
\caption{Upper part: total, point-proton and point-neutron
microscopic LSSM densities of $^{6}$He (from
Ref.~\cite{Karataglidis2000}) (a) and the density of $^{12}$C
\cite{LukyanovZemSlow2004, BurLuk1977} (b). Lower part:
microscopic OP's $V^{DF}$ (c) and $W^{H}$ (d) for the
$^{6}$He+$^{12}$C elastic scattering at $E=3$, 38.3 and 41.6
MeV/nucleon ($N_R=N_I=1$ and $N_I^{sf}=0$).\label{fig1_dens}}
\end{figure}

In Fig.~\ref{fig1_dens} are shown the densities of $^{6}$He and
$^{12}$C, as well as the OP's $V^{DF}$ calculated using
Eqs.~(\ref{eq:1})-(\ref{eq:7}) and $W^{H}$ obtained within the HEA
[Eq.~(\ref{eq:13})] for the three cases of the incident
energy that are considered ($E=3$, 38.3 and 41.6 MeV/nucleon). It can
be seen that the increase of the energy leads to reduced depths and
slopes of ReOP and ImOP.

We calculated the $^{6}$He+$^{12}$C elastic scattering differential
cross sections using the program DWUCK4 \cite{DWUCK} and the
microscopically calculated OP [Eq.~(\ref{eq:14})]. As already
mentioned in Sec.~\ref{s:theory}, in the calculations we add to these
volume potentials the respective surface terms
[Eqs.~(\ref{eq:15})-(\ref{eq:18})]. The latter is done only for the
ImOP, having in mind that due to the breakup channel effects there is
a ``loss of the flux'' from the elastic channel. We note that for the
$^{6}$He+$^{12}$C process there is not a spin-orbit contribution to
the OP in contrast to the $^{6,8}$He$+p$ cases.

In our work we consider the set of the $N_{i}$ coefficients ($N_R$,
$N_I$ and $N^{sf}_I$) as parameters to be found out from the
comparison with the empirical data. We should mention (as it had been
emphasized in our previous works \cite{Lukyanov2007, Lukyanov2009}
for $^{6,8}$He$+p$ scattering) that we do not aim to find a complete
agreement with the data. The introduction of the $N$'s related to the
depths of the different components of the OP's can be considered as a
way to introduce a quantitative measure of the deviations of the
predictions of our method from the reality (e.g., the differences of
$N$'s from unity for given energies).

\begin{figure}
\includegraphics[width=1.0\linewidth]{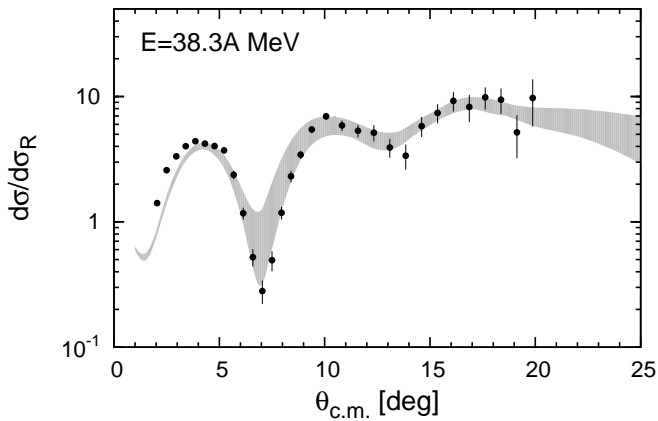}
\caption{Cross sections of $^{6}$He+$^{12}$C elastic scattering at
$E=38.3$ MeV/nucleon calculated by fitting the $N_R,~N_I,~N_I^{sf}$
parameters of the microscopic OP [Eqs.~(\ref{eq:14})-(\ref{eq:18})]
(gray area). The experimental data are taken from
Ref.~\cite{Lapoux2002}.\label{fig2_pot}}
\end{figure}

The starting energy of our calculations was $E=38.3$ MeV/nucleon.
At this energy HEA can be applied as a good approximation to
calculate the ImOP. As a first example, we present in
Fig.~\ref{fig2_pot} the results of our calculations using: i) only
volume OP (\ref{eq:14}) for the two types of ImOP ($V^{DF}$ and
$W^{H}$) and ii) different forms of the surface contributions to
the ImOP [Eqs.~(\ref{eq:15})-(\ref{eq:18})]. The parameters $N$'s
are determined by a fitting procedure. The results of the
calculations are close to each other and that is why all of them
are presented inside a gray area. Two definitions of $\chi^2$ are
used:
\begin{equation}\label{eq:19}
\chi^2 = \frac{1}{N} \sum\limits_{i=1}^{N} \Bigl[
\frac{\sigma^{exp}(\vartheta_i) - \sigma^{th} (\vartheta_i)}{\Delta
\sigma^{exp}(\vartheta_i)} \Bigr]^2,
\end{equation}
\begin{equation}\label{eq:20}
\chi^2_{\sigma} = \frac{1}{N} \sum\limits_{i=1}^{N}
\frac{[\sigma^{exp}(\vartheta_i) -
\sigma^{th}(\vartheta_i)]^2}{\sigma^{th}(\vartheta_i)}.
\end{equation}
In the first definition the $\chi^2$ values were obtained
considering uniform 10$\%$  errors for all the analyzed data. This
procedure is often used by other authors. Here $\sigma^{th}(\vartheta_i)$
and $\sigma^{exp}(\vartheta_i)$ are the theoretical and experimental
values of the differential cross sections ($d\sigma/d\Omega$) or their
ratio to the Rutherford cross section.
In the last case the $\chi^2_\sigma$ is dimensionless and its values
for the results in Fig.~\ref{fig2_pot} range in the interval $0.191
\leq \chi^2_\sigma \leq 0.362$, while the values of $N_R$ are in the
interval $0.893 \leq N_R \leq 1.268$. The values of the predicted
total reaction cross section $\sigma_R$ are also calculated.

One can see from Fig.~\ref{fig2_pot} that the inclusion of different
forms of the surface potential leads to almost similar results for
the cross section. This was also the case of $^{8}$He$+p$ processes
studied in our previous work \cite{Lukyanov2009}. As is known, the
problem of the ambiguity of the values of $N$'s arises when the
fitting procedure concerns a limited number of experimental data.

The situation is ambiguous also for the case of the energy $E=41.6$
MeV/nucleon.

The case of $E=3$ MeV/nucleon is a particular one because of this
rather low energy. Nevertheless, we made an attempt to consider it
using OP obtained in our method. The calculations showed that for
this energy the fitting procedure led to $N^{sf}_I=0$ for the case
of the surface term given by Eq.~(\ref{eq:16}). We note that in
the case of $E=3$ MeV/nucleon the ambiguity in the explanation of
the data \cite{Milin2004} still remains.

In what follows, we tried to choose the most physical values of
$N$'s for the energies considered. As is known, the fitting
procedure belongs to the class of the ill-posed problems (e.g.,
Ref.~\cite{Tikhonov1977}). To resolve this problem it is necessary
to impose some physical constraints when fitting the parameters of
a model. In our case it might be the data on the total cross
sections but often the corresponding values are missing. Another
physical criterion that has to be imposed is the obtained
potentials to obey a determined behavior of the volume integrals
\cite{Satchler79}
\begin{eqnarray}
J_V & = & -\frac{4\pi}{A_p A_t} \int N_R V^{DF}(r) r^2dr,\label{eq:21}\\
J_W & = & -\frac{4\pi}{A_p A_t} \int N_I W(r) r^2dr,\label{eq:22}
\end{eqnarray}
as functions of the energy. Indeed, it was shown for nucleon and
light-ions scattering on nuclei (see, e.g.,
Refs.~\cite{Romanovsky98,Mahaux82,Brandan1997}) that the values of
the volume integrals $J_V$ decrease with the energy increase at
$E<100$ MeV/nucleon, while $J_W$ increases at low energies up to
10-20 MeV/nucleon and then saturates. We would like to note that
such conditions were also imposed in Ref.~\cite{Lukyanov2009} when
the microscopic OP's were introduced to study the $^8$He+p
scattering and their depth parameters $N_R$ and $N_I$ were fitted.
The values of $J_V$ and $J_W$ for the $^6$He+$^{12}$C scattering
that fulfil the condition for their energy dependence are
presented in Tables \ref{tab1}, \ref{tab2} and \ref{tab3}. In the
cases when we include surface terms to the ImOP we modify $J_W$
accounting for them.
\begin{figure}
\includegraphics[width=1.0\linewidth]{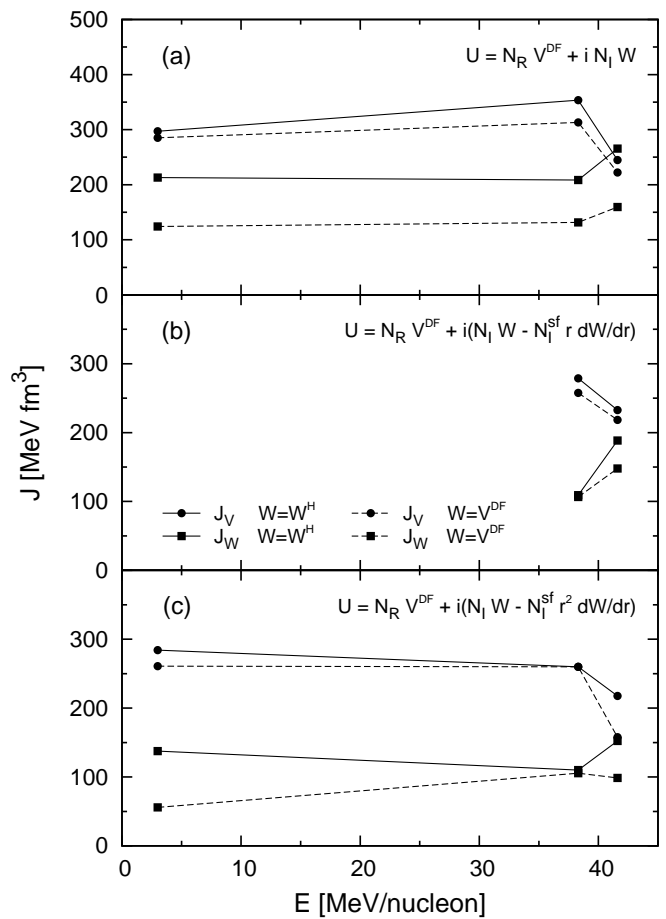}
\caption{The energy dependence of the volume integrals $J_V$ and
$J_W$ that are related to the selected OP's from a number of
fitted microscopic OP's with and without surface terms. The values
of $N$'s, $J_V$, $J_W$, $\chi^2$ and $\sigma_R$ corresponding to
the symbols in panels (a), (b), and (c) are given in
Tables~\ref{tab1}, \ref{tab2}, and \ref{tab3},
respectively.\label{fig3}}
\end{figure}

In the next part of the work we select those sets of the parameters
$N$'s that lead to the already mentioned behavior of $J_V$ and $J_W$
as functions of the energy for three cases: 1) only the volume terms;
2) the volume terms plus the surface term given by Eq.~(\ref{eq:16});
3) the volume terms plus the surface term from Eq.~(\ref{eq:17}). The
behavior of the volume integrals as functions of the energy is
presented in Fig.~\ref{fig3}.

\begin{figure}
\includegraphics[width=1.0\linewidth]{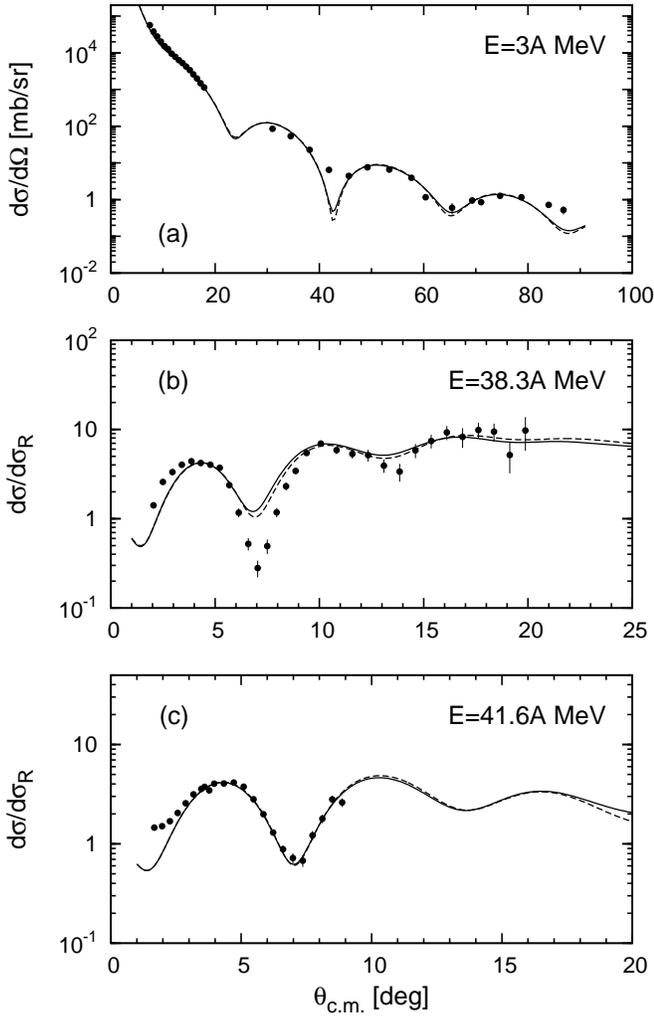}
\caption{Differential cross section of elastic $^{6}$He+$^{12}$C
scattering at $E=3$ (a), 38.3 (b) and 41.6 MeV/nucleon (c)
calculated using only volume OP [Eq.~(\ref{eq:14})]. Solid line:
$W=W^H$, dashed line: $W=V^{DF}$. The values of the fitted
parameters $N_R$ and $N_I$ corresponding to the curves in the
upper, middle and lower part are given in Table \ref{tab1}. The
experimental data are taken from Refs.~\cite{Milin2004,
Lapoux2002, Khalili96}. \label{fig4} }
\end{figure}

\begin{figure}
\includegraphics[width=1.0\linewidth]{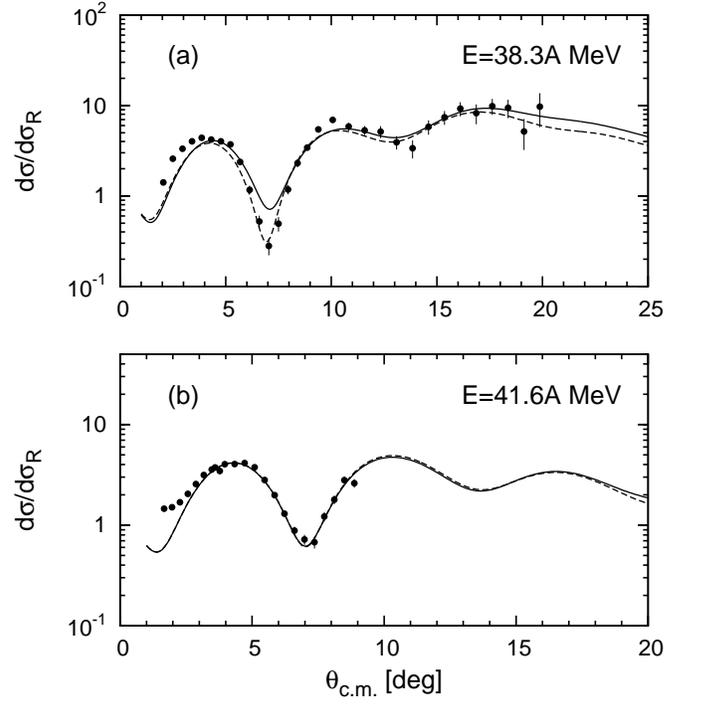}
\caption{The same as in Fig.~\ref{fig4} (without the case for
$E=3$ MeV/nucleon) at $E=38.3$ (a) and $E=41.6$ MeV/nucleon (b)
calculated using volume OP and surface contribution to the ImOP
[Eq.~(\ref{eq:16})]. The values of the parameters $N$'s are given
in Table~\ref{tab2}. The experimental data are taken from
Refs.~\cite{Lapoux2002, Khalili96}.\label{fig5}}
\end{figure}

\begin{figure}
\includegraphics[width=1.0\linewidth]{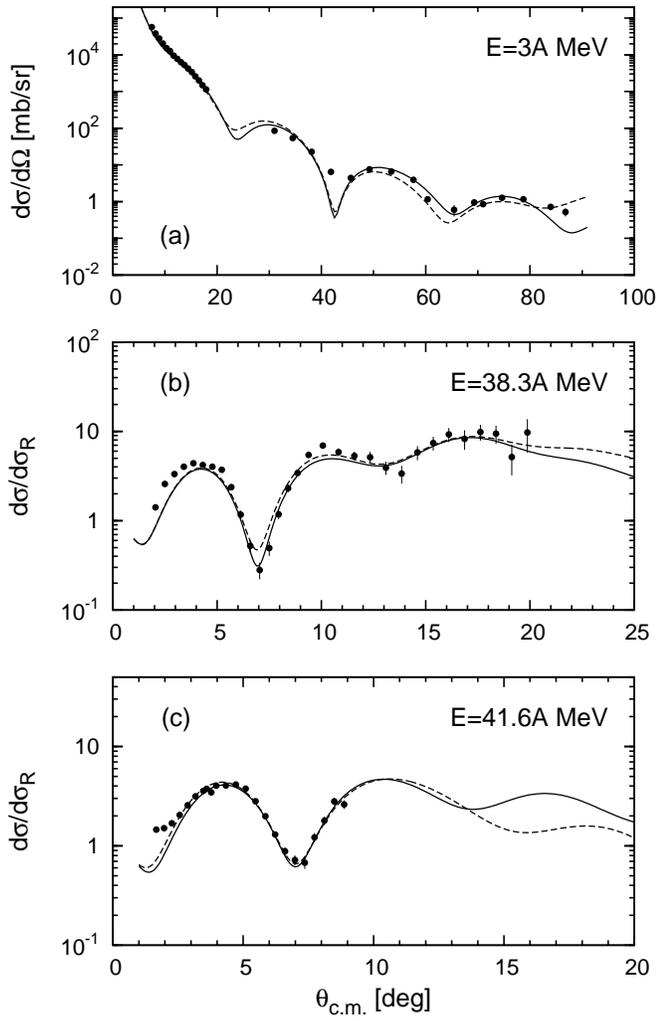}
\caption{The same as in Fig.~4 but for the volume OP and surface
contribution to the ImOP [Eq.~(\ref{eq:17})]. The corresponding
$N$'s are given in Table~\ref{tab3}. The experimental data are
taken from Refs.~\cite{Milin2004, Lapoux2002,
Khalili96}.\label{fig6}}
\end{figure}

Using the same values of the parameters $N$'s already selected, we
present in Figs.~\ref{fig4}, \ref{fig5} and \ref{fig6} the cross
sections for the three energies and for the three cases mentioned
above. One can see that the best agreement with the data for all the
three energies can be obtained by the OP with the volume and surface
term [Eq.~(\ref{eq:17})] whose volume integrals follow (though
approximately) the already mentioned energy dependence.
\begin{table}
\caption{The optimal values of the parameters $N_R$, $N_I$ for the
volume OP [Eq.~(\ref{eq:14})] for the elastic $^6$He+$^{12}$C cross
sections at energies $E=3$, 38.3 and 41.6 MeV/nucleon when the
imaginary potential $W$ was selected in the forms $W^H$ or $V^{DF}$.
The values of the volume integrals $J_V$ and $J_W$, $\chi^2$  and
total reaction cross sections $\sigma_{R}$ (in mb) are also
given.\label{tab1}}
\begin{center}
\begin{tabular}{lccccccc}
\hline\noalign{\smallskip}
$E$ & $W$ & $N_R$ & $N_I$ & $J_V$ & $J_W$ & $\chi^2$ & $\sigma_{R}$ \\
\noalign{\smallskip}\hline\noalign{\smallskip}
3    & $W^H$    & 0.826 & 0.154 & 297.109 & 212.952 & 9.121  & 1427.33 \\
3    & $V^{DF}$ & 0.793 & 0.345 & 285.239 & 124.095 & 9.890  & 1428.52 \\
38.3 & $W^H$    & 1.268 & 0.511 & 353.442 & 208.567 & 80.808 & 1028.77 \\
38.3 & $V^{DF}$ & 1.123 & 0.472 & 313.025 & 131.565 & 50.847 & 1033.79 \\
41.6 & $W^H$    & 0.897 & 0.689 & 244.933 & 265.680 & 3.737  & 1067.32 \\
41.6 & $V^{DF}$ & 0.814 & 0.584 & 222.269 & 159.466 & 3.774  & 1067.55 \\
\noalign{\smallskip} \hline
\end{tabular}
\end{center}
\end{table}
\begin{table}
\caption{The same as Table~\ref{tab1} but for the parameters $N_R$,
$N_I$, and $N_I^{sf}$ of the total OP with the surface term from
Eq.~(\ref{eq:16}).\label{tab2}}
\begin{center}
\begin{tabular}{lcccccccc}
\hline\noalign{\smallskip}
$E$ & $W$ & $N_R$ & $N_I$ & $N_I^{sf}$ & $J_V$ & $J_W$ & $\chi^2$ & $\sigma_{R}$ \\
\noalign{\smallskip}\hline\noalign{\smallskip}
38.3 & $W^H$    & 1.000 & 0.023 & 0.082 & 278.740 & 109.172 & 17.399 & 1055.67 \\
38.3 & $V^{DF}$ & 0.924 & 0.082 & 0.101 & 257.556 & 106.420 & 5.006  & 1174.66 \\
41.6 & $W^H$    & 0.852 & 0.337 & 0.051 & 232.645 & 188.590 & 3.734  & 1070.77 \\
41.6 & $V^{DF}$ & 0.800 & 0.500 & 0.014 & 218.446 & 147.876 & 3.781  & 1072.40 \\
\noalign{\smallskip} \hline
\end{tabular}
\end{center}
\end{table}
\begin{table}
\caption{The same as Table~\ref{tab1} but for the parameters $N_R$, $N_I$, and
$N_I^{sf}$ of the total OP with the surface term from
Eq.~(\ref{eq:17}). The values of $N^{sf}_I$ are in
fm$^{-1}$.\label{tab3}}
\begin{center}
\begin{tabular}{lcccccccc}
\hline\noalign{\smallskip}
$E$ & $W$ & $N_R$ & $N_I$ & $N_I^{sf}$ & $J_V$ & $J_W$ & $\chi^2$ & $\sigma_{R}$ \\
\noalign{\smallskip}\hline\noalign{\smallskip}
3    & $W^{H}$  & 0.790 & 0.074 & 0.002 & 284.160 & 137.506 & 8.912 & 1449.98 \\
3    & $V^{DF}$ & 0.725 & 0.040 & 0.008 & 260.779 & 55.924  & 9.418 & 1533.22 \\
38.3 & $W^H$    & 0.932 & 0.028 & 0.019 & 259.786 & 110.017 & 5.059 & 1185.22 \\
38.3 & $V^{DF}$ & 0.932 & 0.204 & 0.012 & 259.786 & 105.469 & 8.425 & 1161.91 \\
41.6 & $W^H$    & 0.797 & 0.255 & 0.011 & 217.627 & 152.281 & 3.711 & 1091.46 \\
41.6 & $V^{DF}$ & 0.578 & 0.041 & 0.022 & 157.827 &  98.546 & 2.398 & 1224.91 \\
\noalign{\smallskip} \hline
\end{tabular}
\end{center}
\end{table}

\begin{figure}[htb]
\includegraphics[width=1.0\linewidth]{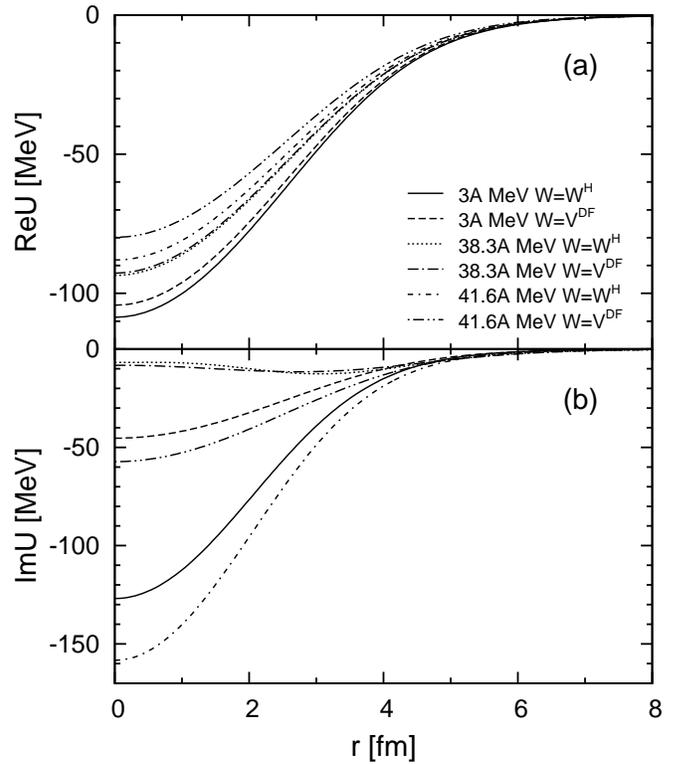}
\caption{Selected OP's that lead to best agreement with the data
of the $^{6}$He+$^{12}$C cross sections shown in Figs.~\ref{fig4},
\ref{fig5}, and \ref{fig6}. The values of the parameters $N$'s for
the curves at $E=3$ and 41.6 MeV/nucleon (with $W=W^H$ and
$W=V^{DF}$) are listed in Table~\ref{tab1}, while at $E=38.3$
MeV/nucleon are given in Table~\ref{tab2} for $W=V^{DF}$ and
Table~\ref{tab3} for $W=W^H$.\label{fig7}}
\end{figure}

In Figs.~\ref{fig7}(a) and \ref{fig7}(b) are given those
microscopic ReOP and ImOP that lead to the best agreement with the
experimental data for the $^{6}$He+$^{12}$C elastic scattering
cross sections shown in Figs.~\ref{fig4}, \ref{fig5} and
\ref{fig6} for energies $E=3$, 38.3 and 41.6 MeV/nucleon. For
$E=3$ MeV/nucleon there are only volume OP's, while for $E=38.3$
MeV/nucleon there is a surface contribution to the ImOP using
$W=V^{DF}$ in Eq.~(\ref{eq:16}) and $W=W^H$ in Eq.~(\ref{eq:17});
for $E=41.6$ MeV/nucleon there are only volume OP's.

\begin{figure}[htb]
\includegraphics[width=1.0\linewidth]{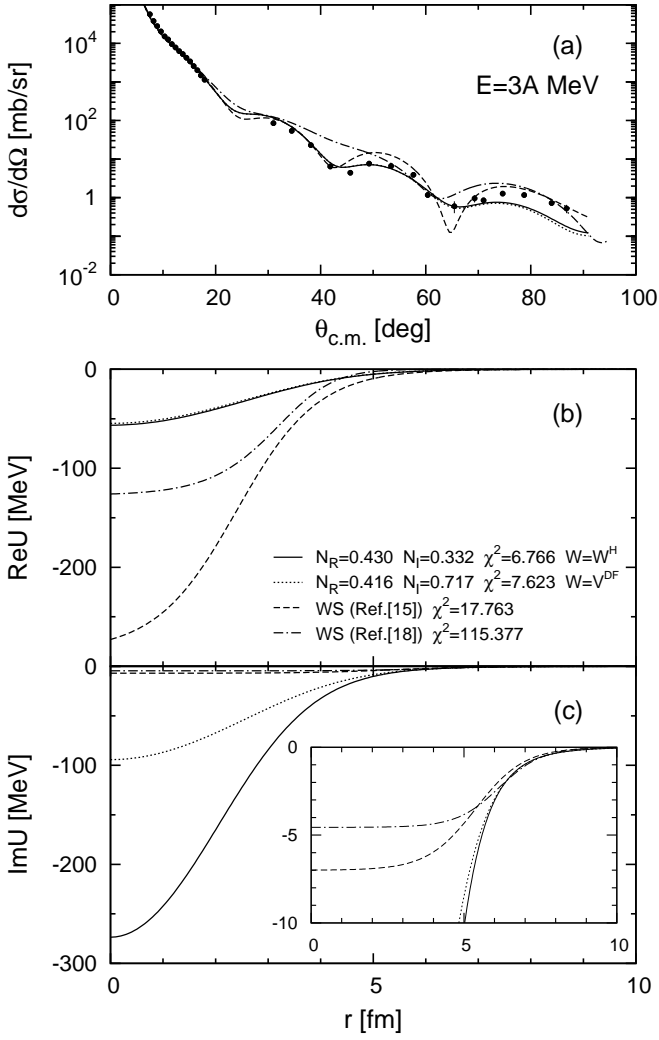}
\caption{(a) Differential cross section of elastic
$^{6}$He+$^{12}$C scattering at $E=3$ MeV/nucleon. Solid and
dotted lines show the results with microscopic ImOP $W^{H}$ and
$V^{DF}$, respectively. The results with the phenomenological OP's
from Refs.~\cite{Milin2004} and \cite{Kucuk2009} are given by
dashed and dash-dotted lines, correspondingly. The experimental
data are taken from Ref.~\cite{Milin2004}; The ReOP and ImOP for
$E=3$ MeV/nucleon microscopically obtained in the present work and
those from Refs.~\cite{Milin2004} and \cite{Kucuk2009} are given
in panels (b) and (c), respectively. \label{fig8}}
\end{figure}

In Figs.~\ref{fig8}(b) and \ref{fig8}(c) the real and imaginary
parts of the OP for $E=3$ MeV/nucleon obtained in the present work
are compared with the phenomenological OP's from
Ref.~\cite{Milin2004} (where WS forms have been used for ReOP and
ImOP) and from Ref.~\cite{Kucuk2009} (with OP having a squared WS
real part and a standard WS shape for the ImOP). The results for
the cross sections are shown in Fig.~\ref{fig8}(a). One can see
much better agreement for our cross sections obtained using
microscopic OP's than those obtained in a phenomenological way in
Refs.~\cite{Milin2004,Kucuk2009}.

We should note also that in Ref.~\cite{Kucuk2009} the ReOP ($V_0$)
increases and the ImOP ($W_0$) decreases with the energy increase
which is in contradiction with the generally accepted results and
with the behavior of the volume integrals as functions of the
energy \cite{Romanovsky98,Mahaux82,Brandan1997}.

\begin{figure}[htb]
\includegraphics[width=1.0\linewidth]{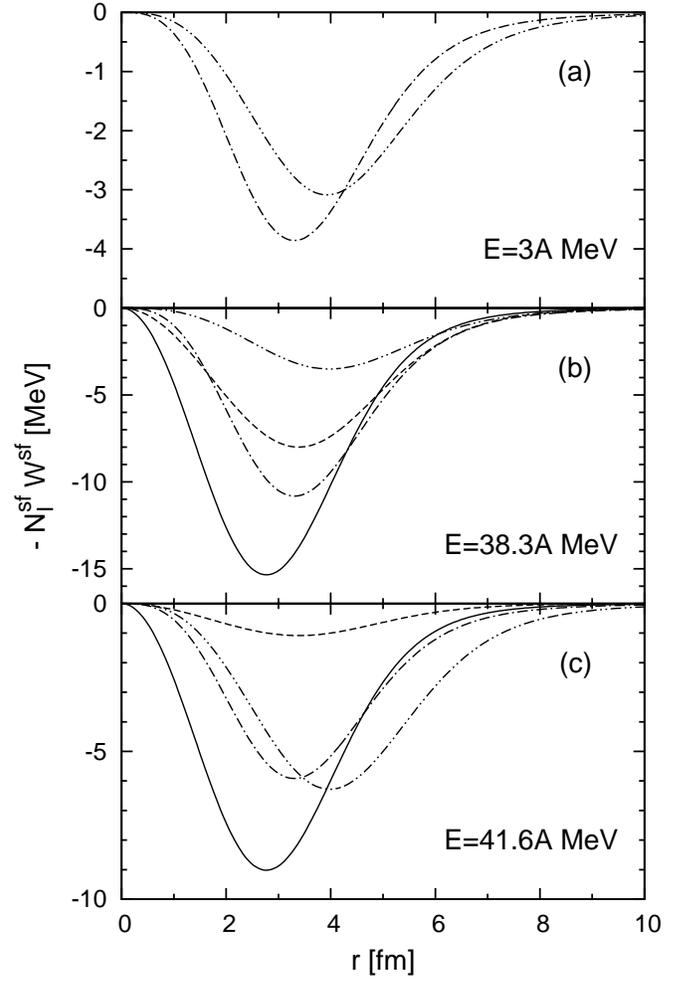}
\caption{The surface ImOP's (dynamical polarization OP's) used in
the calculations of the cross sections of $^{6}$He+$^{12}$C
elastic scattering at $E=3$ (a), 38.3 (b), and 41.6 MeV/nucleon
(c). Solid and dashed lines: using Eq.~(\ref{eq:16}) with values
of $N$'s from Table~\ref{tab2} and with $W=W^H$ and $W=V^{DF}$,
respectively. Dot-dashed and dashed two-dots lines: using
Eq.~(\ref{eq:17}) with $N$'s from Table~\ref{tab3} and with
$W=W^H$ and $W=V^{DF}$, respectively. \label{fig9}}
\end{figure}

In Fig.~\ref{fig9} the deviations of the OP's from their volume
parts are presented for the three energies considered. In our
theoretical scheme they are related to the surface parts of the
ImOP in the form of $N_I^{sf} r dW^{H(DF)}/dr$ and $N_I^{sf} r^{2}
dW^{(H)DF}/dr$. As is known, these contributions can be considered
as the so-called dynamical polarization potential that owes its
origin to effects of the breakup of a pair of neutrons from
$^6$He.

In what follows, we would like to discuss the deviations of the
values of the coefficients $N$'s (and, particularly, of $N_{I}$)
from unity. As known, a folding model can be thought as meaningful
only when the renormalization coefficients of the folded potential
are close to unity. In our case (see Table~\ref{tab1})
$N_{I}=0.154$ for $E=3$ MeV/nucleon and $N_{I}=0.689$ for $E=41.6$
MeV/nucleon. Here we would like to emphasize that the obtained
values of $N_{I}$ within the HEA for the small energy $E=3$
MeV/nucleon reflect the effects of Pauli blocking (see, e.g.
\cite{Hussein1991}) that in the case of nucleus-nucleus scattering
reduce the depth of ImOP by a factor of 10 or more. This is
related to the fact that in the HEA the microscopic optical
potential is proportional to the free $NN$ cross section
($\sigma_{N}$), that in nuclear matter is reduced by the so-called
in-medium factor $f_{m}$ that accounts for, in particular, the
Pauli blocking effect:
\begin{equation}\label{eq:23}
\sigma_N^{(m)}=\sigma_N f_m.
\end{equation}
There are many estimations of the factor $f_{m}$, e.g. those based
in the Br\"{u}ckner-Hartree-Fock (BHF) theory
\cite{Lee1993,Xiangzhou1998,Bertulani2010}. In
\cite{Xiangzhou1998} the expression for $f_{m}$ is obtained by a
least squares fit to the experimental total reaction cross section
data over a wide incident energy range. The latter gives the
separate forms for the $\sigma_{pp}$ and $\sigma_{pn}$ cross
sections accompanied by the factors
\begin{equation}\label{eq:24}
f_m(np)=\frac{1+20.88\,\,E^{\,0.04}\,\,\rho^{\,2.02}}
{1+35.86\,\,\rho^{1.90}},
\end{equation}
\begin{equation}\label{eq:25}
f_m(nn)=\frac{1+7.772\,\,E^{\,0.06}\,\,\rho^{\,1.48}}
{1+18.01\,\,\rho^{\,1.46}},
\end{equation}
where $E$ is the kinetic energy in laboratory system per nucleon
of the projectile nucleus, and $\rho=\rho_p+\rho_t$. (One can
guess that in Eqs.~(\ref{eq:24}) and (\ref{eq:25}) the numerical
coefficients in the second terms of $f_m$ have the respective
dimensions that to measure $E_{lab}$ in MeV and $\rho$ in
fm$^{-3}$). Recently, expressions for $f_{m}$ were presented in
Ref.~\cite{Bertulani2010} in the approximation of the isotropic
free $NN$ cross section. In this approach the Pauli projection
operator in the $G$-matrix as a solution of the Bethe-Goldstone
equation is considered as a geometrical factor that restricts the
conditions on the available angles of scattering of the $NN$ pair
to unoccupied final states. The result for the $NN$ correction
factor is in the form \cite{Bertulani2010}:
\begin{equation}\label{eq:26}
f_m=\frac{1}
{1+1.892\left(\frac{|\rho_p-\rho_t|}{\bar\rho\rho_0}\right)^{2.75}}
f(E),
\end{equation}
where
\begin{equation}\label{eq:27}
f(E>46.27{\bar\rho}^{2/3})= 1-{37.02\over E},
\end{equation}
\begin{equation}\label{eq:28}
f(E<46.27{\bar\rho}^{2/3})={E\over 231.38{\bar\rho}^{2/3}},
\end{equation}
$E$ is the laboratory energy per nucleon in MeV, $\rho_0=0.17$
fm$^{-3}$ and $\bar\rho=(\rho_p+\rho_t)/\rho_0$. In practice, hard
numerical problems arise when one intends to use these formulae in
calculations of folding integrals for the microscopic potentials.
Instead, it is easy to estimate the in-medium effects in the
realistic case suggesting that the main contribution comes from
the region of half-density radii of the colliding nuclei, where
$\rho=\rho_p+\rho_t=\rho_0$, $\rho_p-\rho_t=0$ and ${\bar\rho}=1$.
Then, the first term in Eq.~(\ref{eq:26}) equals to 1, and thus
one gets $f(E)\simeq$ 0.013 for $E=3$ MeV/nucleon and $f(E)\simeq$
0.18 for $E=41.6$ MeV/nucleon. The use of Eqs.~(\ref{eq:24}) and
(\ref{eq:25}) (from \cite{Xiangzhou1998}) lead to the following
estimations: $f_{m}(np)=0.717$ and $f_{m}(nn)=0.68$ for $E=3$
MeV/nucleon and $f_{m}(np)=0.75$ and $f_{m}(nn)=0.74$ for $E=41.6$
MeV/nucleon. One can see that the results for $N_{I}=0.154$ (at
$E=3$ MeV/nucleon) and $N_{I}=0.689$ (at $E=41.6$ MeV/nucleon)
from our work mentioned above and obtained by a fitting procedure
are in a correct "direction" and they are between the estimations
using Refs.~\cite{Xiangzhou1998} and \cite{Bertulani2010}.

We note that these estimations are valid only for the volume OP's
(even in the local density approximation). We emphasize that one
has to account also for the competition with the channels at the
nuclear periphery (the breakup) that, according to the
coupled-channel calculations, play an important role. Their
contribution that had been initially obtained only accounting for
the channels of $NN$ scattering inside the nuclear matter leads to
changes of the ImOP in the elastic channel. An example for such a
"renormalization" of ImOP for elastic $d+A$ scattering due to the
stripping channel was given in Ref.~\cite{Ingemarsson1999}. For
more complex systems this is difficult to be done because only in
the $(d,p)$ reaction one can use the approximation of the
"delta"-potential in the deuteron. Thus, we note that not only the
Pauli blocking, but also other processes due to different
mechanisms (breakup and others) play role, as it is considered in
various references mentioned in the Introduction of the present
work.

\section{Summary and conclusions\label{s:conclusions}}

The results of the present work can be summarized as follows:

(i) The microscopic optical potential and cross sections of
$^{6}$He+$^{12}$C elastic scattering were calculated at the
energies of $E=3$, 38.3 and 41.6 MeV/nucleon. Comparisons with the
experimental data and results of other approaches were presented.
The direct and exchange parts of the real OP ($V^{DF}$) were
calculated microscopically using the double-folding procedure and
density-dependent M3Y (of CDM3Y6-type) effective interaction based
on the Paris nucleon-nucleon potential. The imaginary parts of the
OP were taken in the forms of $V^{DF}$ or $W^H$, the latter being
calculated using the high-energy approximation. The microscopic
densities of protons and neutrons in $^{6}$He calculated within
the large-scale shell model were used. The nucleon density
distribution functions of $^{12}$C were taken as defolded charged
densities obtained from the best fit to the experimental form
factors from electron elastic scattering on $^{12}$C. In this way,
in contrast to the phenomenological and semi-microscopic models we
deal with fully microscopic approach as a physical ground to
account for the single-particle structure of the colliding nuclei.
The elastic scattering differential cross sections were calculated
using the program DWUCK4.

(ii) While at low energies the volume OP's can reproduce sufficiently
well the experimental data, at higher energies additional surface
terms in OP having a form of a derivative of the imaginary part of
the OP became necessary and were used in the present work.

(iii) The depths of the real and imaginary parts of the microscopic
OP's are considered as fitting parameters. As is expected when one
utilizes the fitting procedure in the case of a limited number of
experimental data, the problem of the ambiguity of these parameters
arises. To overcome (at least partly) this ambiguity, additional
physical constraints should be imposed. Doing so, we require in our
work the values of the depth's parameters $N$'s to lead to  volume
integrals $J_V$ and $J_W$ with realistic energy dependence for
energies $0 < E < 100$ MeV/nucleon. Namely, $J_V$'s must decrease
while $J_W$'s increase to some constant values with the increase of
the energy.

(iv) The comparison of our results with those of some
phenomenological approaches pointed out the advantages of using
microscopic real and imaginary parts of the optical potential
imposing realistic physical constraints on their depths as that one
of the behavior of the volume integrals as functions of the energy.

(v) As in works of other authors (e.g., Ref.~\cite{Khoa1993}) we
consider in more details the behavior of the OP in the nuclear
periphery. This gives a possibility to make some conclusions about
the contributions of the dynamical polarization terms of the OP or,
in other words, about the coupled-channel effects.

(vi) It is shown that the deviations of the values of $N$'s from
unity (given in Table~\ref{tab1} for the volume OP's) that are
obtained by a fit to the experimental data for $^{6}$He+$^{12}$C
elastic scattering are related to the Pauli blocking effects.
These values are smaller than unity and turn out to be between the
approximate estimations performed using the results of approaches
where Pauli blocking is taken into account (e.g.
Refs.~\cite{Xiangzhou1998,Bertulani2010}). It is also mentioned
that together with the important Pauli blocking effects, the role
of other mechanisms, such as breakup processes, also have to be
accounted for in the study of the reaction considered.

\begin{acknowledgments}
The work is partly supported by the Project from the Agreement for
co-operation between the INRNE-BAS (Sofia) and JINR (Dubna). Three of
the authors (D.N.K., A.N.A. and M.K.G.) are grateful for the support
of the Bulgarian Science Fund under Contract No.~02--285 and one of
them (D.N.K.) under Contract No.~DID--02/16--17.12.2009. The authors
E.V.Z. and K.V.L. thank the Russian Foundation for Basic Research
(Grant No. 09-01-00770) for the partial support.
\end{acknowledgments}

\end{document}